\def\refnum{\addtocounter{enumi}{1}\arabic{enumi}}
\def\rn{(\refnum)~}
\def\nn#1 #2{#2. #1}				
\def\nnn#1 #2 #3{#2. #3. #1}			
\def\nnnn#1 #2 #3 #4{#2. #3. #4 #1}		
\def\nnnnn#1 #2 #3 #4 #5{#2. #3. #4 #5. #1}	
\def\dualand{ and\hbox{ }}				
\def\multiand{, and\hbox{ }}				
\def\rf#1;#2;#3;#4;#5 {{\frenchspacing\par\rn#1, #3 {\bf #4}, #5 (#2). \par}}
\def\rg#1;#2;#3;#4;#5;#6 {{\frenchspacing\par\rn#1, #3 {\bf #4}, #5 (#2). \par}}
\def\rfbook#1;#2;#3;#4;#5 {{\frenchspacing\par\rn#1, {\it #3} (#5, #4, #2).\par}}
\def\rfprep#1;#2;#3 {{\par\frenchspacing\rn#1, #3 (#2).\par}}
\def\preskip {\vskip-0.0cm}

\def\beq#1{\begin{equation}\label{#1}}
\def\eeq{\end{equation}}
\def\beqa#1{\begin{eqnarray}\label{#1}}
\def\eeqa{\end{eqnarray}}

\def\spose#1{\hbox to 0pt{#1\hss}}
\def\simlt{\mathrel{\spose{\lower 3pt\hbox{$\mathchar"218$}} \raise 2.0pt\hbox{$\mathchar"13C$}}}
\def\simgt{\mathrel{\spose{\lower 3pt\hbox{$\mathchar"218$}} \raise 2.0pt\hbox{$\mathchar"13E$}}}
\def\simpropto{\mathrel{\spose{\lower 3pt\hbox{$\mathchar"218$}} \raise 2.0pt\hbox{$\propto$}}}

\def\bt{\begin{tabbing}}
\def\et{\end{tabbing}}
\def\beq#1{\begin{equation}\label{#1}}
\def\eeq{\end{equation}}

\def\sec#1{Section~\ref{#1}}

\def\bfig{\begin{figure}[h] \centerline{\hbox{}}\vfill}
\def\efig{\end{figure}\vfill\newpage}

\def\fig#1{Figure~\ref{#1}}

\def\fig#1{Figure~\ref{#1}}
\def\Fig#1{Figure~\ref{#1}}
\def\tabela#1{Table~\ref{#1}}

\def\expec#1{\langle#1\rangle}

\def\l{\ell}
\def\etal{{\frenchspacing\it et al.}}
\def\ie  {{\frenchspacing\it i.e.}}
\def\eg  {{\frenchspacing\it e.g.}}

\def\tab{\hspace{\parindent}}

\def\ns{n_s}
\def\nt{n_t}
\def\As{A_s}
\def\At{A_t}
\def\fn{f_\nu}
\def\Ok{\Omega_{\rm k}  }
\def\Ol{\Omega_\Lambda  }
\def\Oc{\Omega_{\rm cdm}}
\def\On{\Omega_\nu      }
\def\Ob{\Omega_{\rm b}  }
\def\Od{\Omega_{\rm dm} }
\def\ob{\omega_{\rm b}  }
\def\od{\omega_{\rm dm} }

\documentclass[11pt,twoside]{article}
\usepackage{asp2004}
\usepackage{psfig}
\usepackage{epsf}
\usepackage{graphics}
\usepackage{lscape}
\usepackage{color}

\def\edcomment#1{\iffalse\marginpar{\raggedright\sl#1\/}\else\relax\fi}
\reversemarginpar

\begin{document}
\title{The Cosmic Microwave Background and Its Polarization}
\author{Ang\'elica de Oliveira-Costa}
\affil{University of Pennsylvania, Dept. of Physics \& Astronomy, Philadelphia, 19104 PA, USA}

\begin{abstract}
The DASI discovery of CMB polarization,  confirmed by WMAP, 
has opened a new chapter in cosmology. Most of the useful information 
about inflationary gravitational waves and reionization is on large 
angular scales where Galactic foreground contamination is the worst.
The goal of the present review is to provide the state-of-the-art of the CMB 
polarization from a practical point of view, connecting real-world data to 
physical models. We present the physics of this polarized phenomena and 
illustrate how it depends of various cosmological parameters for standard 
adiabatic models. We also present all observational constraints to date and
discuss how much we have learned about polarized foregrounds so far from the 
CMB studies. Finally, we comment on future prospects for the 
measurement of CMB polarization.
\end{abstract}
\thispagestyle{plain}


{\vskip-3.0cm} 

\section{Introduction}\label{intro}

\tab
The recent discovery of cosmic microwave background (CMB) polarization by 
the DASI (Degree Angular Scale Interferometer) experiment \cite{kovac02}, 
confirmed by the WMAP (Wilkinson Anisotropy Probe) satellite 
\cite{kogut03}, has opened a new chapter in cosmology -- see \fig{summary}.
Although CMB polarization on degree scales and below can sharpen cosmological 
constraints and provide an important cross-check on our basic assumptions about 
the behavior of fluctuations in the universe \cite{ZSS97,Eisenstein98}, 
the potential for the most dramatic improvements lies on the largest angular 
scales where it provides a unique probe of the reionization epoch and primordial 
gravitational waves.

CMB polarization is produced via Thomson scattering which occurs either at 
decoupling or during reionization. The level of this polarization is linked 
to the local quadrupole anisotropy of the incident radiation on the scattering 
electrons, and it is expected to be of order 1\%-10\% of the amplitude of the 
temperature anisotropies depending on the angular scale -- see \cite{ZH95,HW97} 
and references therein.

The goal of the present review is to survey the state-of-the-art of  CMB 
polarization from a practical point of view, connecting real-world data to 
physical models. 
In $\sec{physics}$ and $\sec{theo}$, we summarize the physics of CMB polarization
and illustrate how it depends on various cosmological 
parameters for standard adiabatic models. In $\sec{measurements}$, we present 
all observational constraints to date. In $\sec{foregrounds}$, we discuss 
how such measurements can be compromised by the Galactic foregrounds. Finally, 
we conclude in $\sec{conclusions}$ with comments on future prospects for the 
measurement of CMB polarization.


\section{How does CMB polarization form?}\label{physics}

\tab
At times before decoupling, the universe was hot enough that protons and electrons 
existed freely in a plasma. Through Thomson scattering, they kept tightly coupled, 
\ie, they were in thermal equilibrium at a common temperature. As a consequence 
of this tight coupling epoch, the radiation field could only possess a monopole
(given by the plasma's temperature) and a dipole (described as a Doppler shift 
due to the peculiar velocities in the fluid). Any higher moment was damped away
due to scattering, and no net polarization could be produced (\ie, the radiation 
field was unpolarized).
  
As the temperature of the universe drooped (below 3,000K), protons and electrons 
started to recombine into neutral hydrogen. The mean free path grown rapidly and 
the eletrons began to see local quadrupoles within the plasma. At this point, 
Thomsom scattering started to produce polarized light. 
By the time almost all free electrons were used up to produce neutral hydrogen, 
Thomson scattering ceased for lack of scatterers, and the radiation was said to decouple. 
From that point on, this radiation, known as the CMB, propagated freely until the universe 
reionized around $z \approx 17$.


\section{Polarization phenomenology}\label{theo}

\tab
Whereas most astronomers use the Stokes parameters $Q$ and $U$ 
to describe polarization measurements, the CMB community uses 
two scalar fields $E$ and $B$ that are independent of how the 
coordinate system is oriented, and are related to the tensor 
field $(Q,U)$ by a non--local transformation \cite{K97,ZS97,Z98}.
Scalar CMB fluctuations have been shown to generate only curl-free 
$E$-modes, whereas gravity waves, CMB lensing and foregrounds 
generate both $E$ and a pure curl component called $B$-modes\footnote{
	The $B$-type modes exhibit linear polarization at $\pm 45\deg$ 
	to the direction of the polarization gradient.}. 


\subsection{The six power spectra}\label{SixSec} 

\tab
Since CMB measurements can be decomposed into three maps ($T$,$E$,$B$), 
where $T$ denotes the unpolarized component, there are a total of 6 
angular power spectra that can be measured. Expanding the $T$, $E$ and $B$ 
maps in spherical harmonics with coefficients $a_{\l m}^T$, $a_{\l m}^E$ 
and $a_{\l m}^B$, these 6 spectra are defined by
\beqa{3spectra1}
  C_\l^T = \expec{a_{\l m}^{T*} a_{\l m}^T}, \nonumber 
  C_\l^E = \expec{a_{\l m}^{E*} a_{\l m}^E}, \nonumber 
  C_\l^B = \expec{a_{\l m}^{B*} a_{\l m}^B}, \nonumber 
  C_\l^X = \expec{a_{\l m}^{T*} a_{\l m}^E}, \nonumber \\ 
  C_\l^Y = \expec{a_{\l m}^{T*} a_{\l m}^B}, \nonumber 
  C_\l^Z = \expec{a_{\l m}^{E*} a_{\l m}^B}, \nonumber 
\eeqa 
corresponding to 
	$TT$,  
	$EE$,  
	$BB$,  
	$TE$,  
	$TB$ and  
	$EB$  
correlations\footnote{
	From here on, we adopt the notation
	$TT \equiv T$, 
	$EE \equiv E$,
	$BB \equiv B$, 
	$TE \equiv X$, 
	$TB \equiv Y$,
	$EB \equiv Z$. 
	},
respectively \cite{TC01}. 
By parity, $C_\l^Y=C_\l^Z=0$ for scalar CMB fluctuations, but it is nonetheless 
worthwhile to measure these power spectra as probes of both exotic physics 
\cite{KK99,XK99,Kamionkowski00} and foreground contamination \cite{angel_polar}. 
$C_\l^B=0$ for scalar CMB fluctuations to first order in perturbation theory 
\cite{K97,ZS97,Z98,HuWhite97} --- secondary effects such as gravitational 
lensing can create $B$ polarization even if there are only density perturbations 
present \cite{ZSLENS}. 
In the absence of reionization, $C_\l^E$ is typically a couple of orders of 
magnitude below $C_\l^T$ on small scales and approaches zero on the very largest 
scales. 


\subsection{Covariance versus correlation}

\begin{figure}[tb]
\preskip
\centerline{\epsfxsize=6.7cm\epsffile{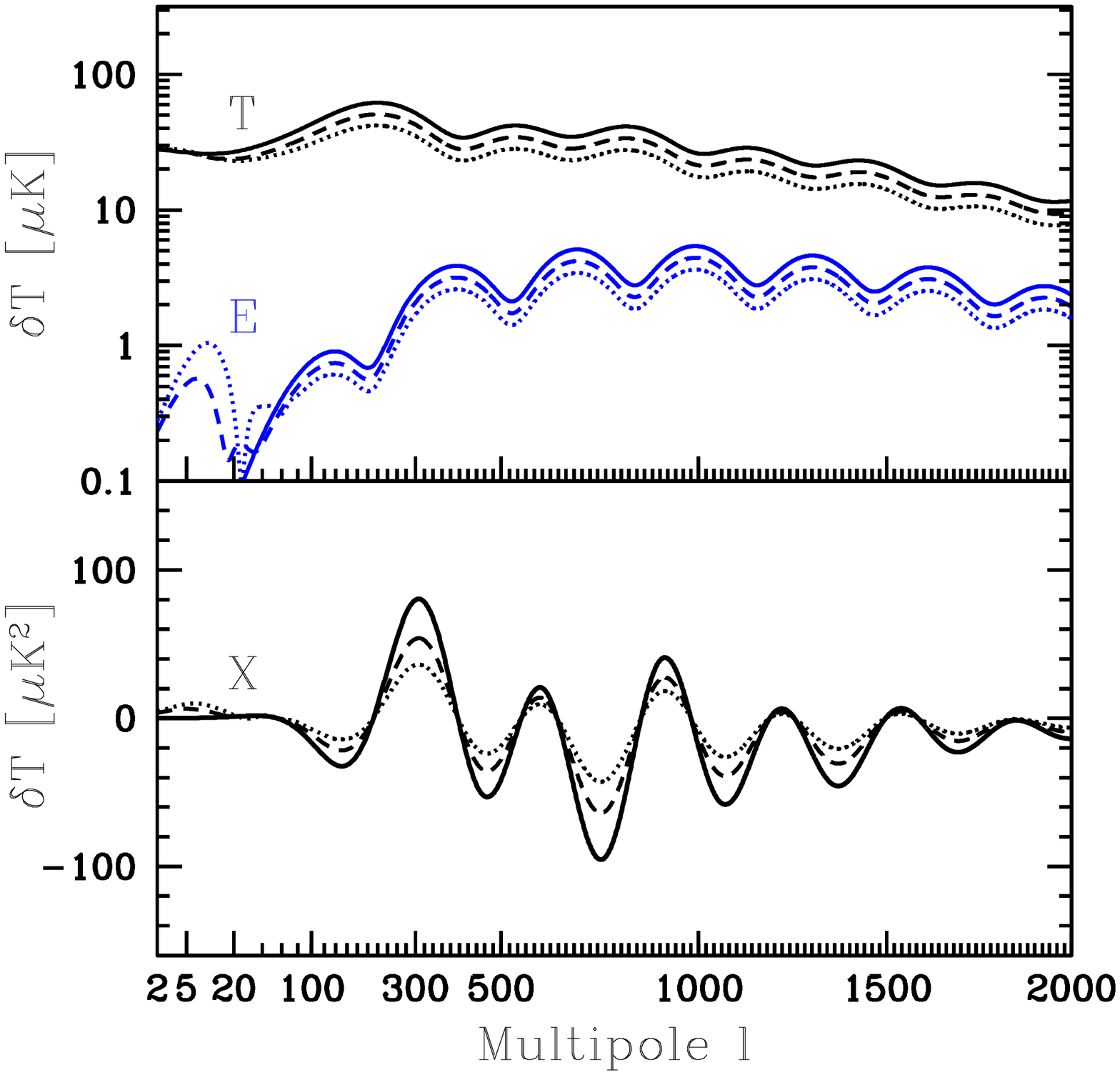} \epsfxsize=6.7cm\epsffile{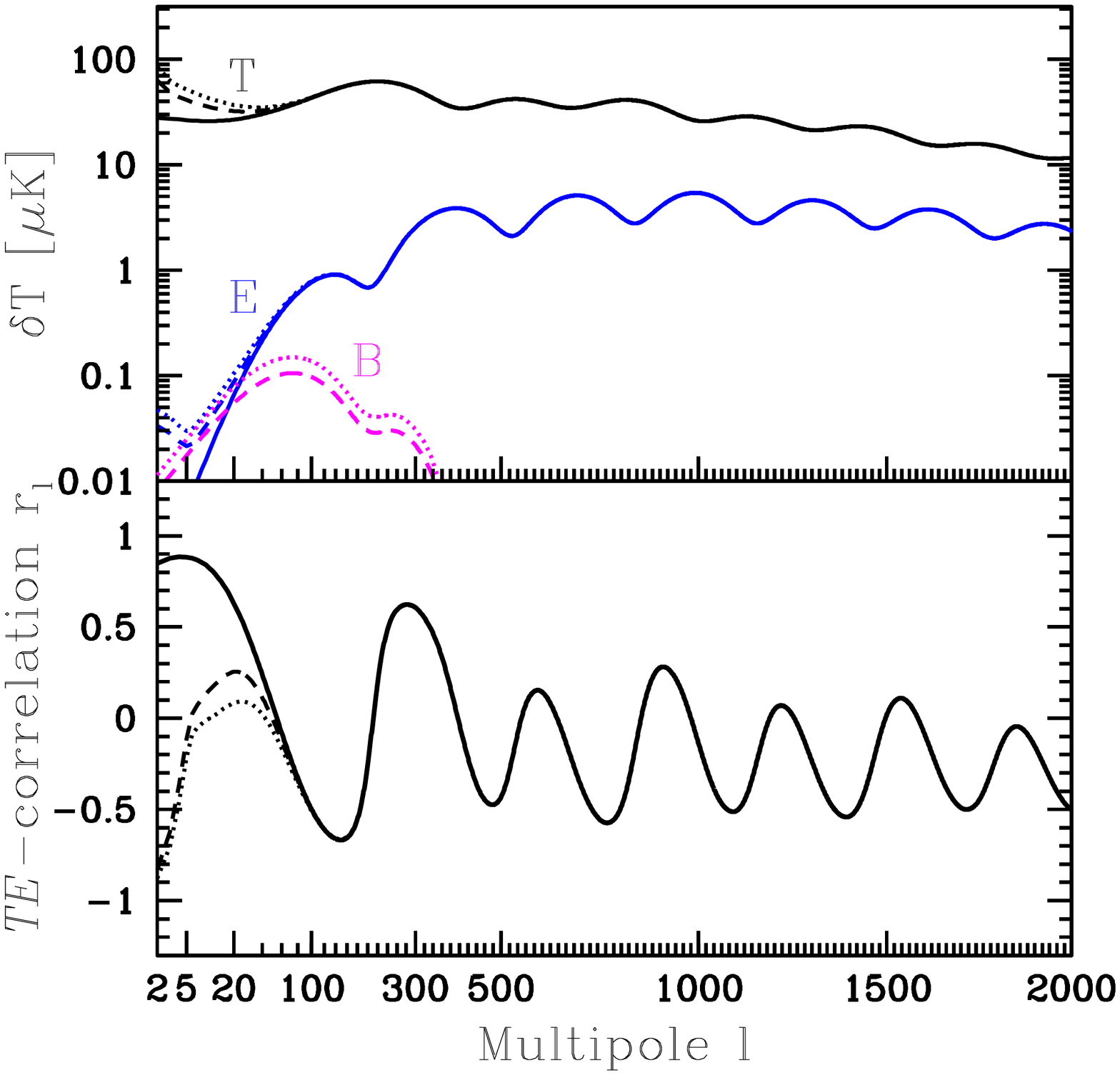}}
\caption{\label{fig1}\footnotesize%
        {\it Left:}
	How the reionization optical depth $\tau$ affects the $T$ and $E$ power 
	spectra (top panel) and the $X$ power spectra (lower panel). Solid, dashed 
	and dotted curves correspond to for $\tau$=0, 0.2 and 0.4, respectively.
	{\it Right:}
         How the gravity wave amplitude $\At$ affects the $T$, $E$ and $B$ 
	power spectra (top panel) and the correlation coefficient $r_\l$ (lower panel). 
	Solid, dashed and dotted curves correspond to for $\At$=0, 0.2 and 0.4, respectively.
        }
\end{figure}

\tab
The cross-power spectrum $C_\l^X$ is not well suited for the usual logarithmic 
power spectrum plot, since it is negative for about half of all $\l$-values 
\cite{angel_pique}. Sometimes, a theoretically more convenient quantity is the 
dimensionless correlation coefficient
	$r^X_\l\equiv {C_\l^X\over (C_\l^T C_\l^E)^{1/2}}$,
plotted on a linear scale in \fig{fig1} (Right,  lower panel), since the 
Schwarz inequality restricts it to lie in the range $-1\le r^X_\l\le 1$. From here 
on we use $r_\l$ as shorthand for $r^X_\l$. For more details about $r_\l$ and 
how it depends on cosmological parameters, see section II.b in \cite{angel_pique}.


\subsection{Cosmological parameter dependence of polarization spectra}
\label{ParameterSec}

\tab
A detailed review of how CMB polarization reflects underlying physical 
processes in given in \cite{HW97}. In this subsection, we briefly review 
this topic from a more phenomenological point of view (see also 
\cite{Jaffe00}), focusing on how different cosmological parameters 
affect various features in the $E$, $B$ and $X$ power spectra. For more 
details, the reader is referred to the polarization movies at
 	{\it www.hep.upenn.edu/$\sim$angelica/polarization.html}.

Let us consider adiabatic inflationary models specified by the following 10 
parameters:
	the reionization optical depth $\tau$, 
	the primordial amplitudes $\As$, $\At$ and 
	tilts $\ns$, $\nt$ of scalar and tensor fluctuations, 
and five parameters specifying the cosmic matter budget. The various 
contributions $\Omega_i$ to critical density are for
	curvature $\Ok$, 
	vacuum energy $\Ol$, 
	cold dark matter $\Oc$, 
	hot dark matter (neutrinos) $\On$ and 
	baryons $\Ob$.
The quantities $\ob\equiv h^2\Ob$ and $\od\equiv h^2\Od$ correspond to the 
physical densities of baryons and total (cold + hot) dark matter 
($\Od\equiv\Oc+\On$), and $\fn\equiv\On/\Od$ is the fraction of the dark 
matter that is hot. The baseline values of the parameters here and in the 
movies are for the concordance model of \cite{X01,spergel03,sdsswmap03}.
All power spectra were computed with the CMBfast software \cite{SZ96}.


\subsubsection{Polarized versus unpolarized:}

If recombination were instantaneous (with the radiation field locally isotropic),
there would be no polarization at all.

Both the $E$ and the $T$ power spectra carry information about the $z\simgt 10^3$ 
pre-recombination epoch in the form of acoustic oscillations. From a practical 
point of view, there are two obvious differences between the $E$ and $T$ power 
spectra as illustrated by \fig{fig1}:
\begin{itemize}
\item The $E$ power is smaller since the polarization percentage is small, 
      making measurements more challenging. This is because polarization 
      is only generated when locally anisotropic radiation scatters off of free 
      electrons, and this only occurs during the brief period when recombination 
      is taking place: before recombination, radiation is quite isotropic and 
      after recombination there is almost no scattering.
\item Aside from reionization effects, the $E$ power approaches zero on scales 
      much larger than those of the first acoustic peak. 
      This is because the polarization anisotropies are only generated on
      scales of order  the mean free path at recombination and below.
\end{itemize}

As detailed below, changing the cosmological parameters affects the polarized 
and unpolarized power spectra rather similarly except for the cases of reionization 
and gravity waves. 
Another interesting difference between the power spectra of  the temperature 
and polarization is that they exhibit peaks which are approximately a half-cycle 
out of phase -- see \fig{fig1}.  
As described above, as recombination proceeds, the eletrons begin to see 
radiation Doppler-shifted by the velocity fields in the plasma and scattering 
leads to polarization.  Since $E$-mode polarization arises from velocities, 
when the  fluid velocity drops to zero, the amplitude of the polarization will fall
to a minimum at the compression or expansion maxima of the density mode. 
Similarly, the amplitude of the polarization will be highest at the density nulls, 
when the fluid velocity reaches a maximum. 
The $X$ spectrum, on the other hand,  has a more complex behavior with the 
sign of the correlation depending on whether the amplitude of the mode was 
increasing or decreasing at the time of decoupling. 


\subsubsection{Reionization}

Reionization at redshift $z_*$ introduces a new scale $\l_*\sim 20 (z_*/10)^{1/2}$ 
corresponding to the horizon size at the time. Primary (from $z\simgt 1000$) 
fluctuations $\delta T_\l$ on scales $\l\gg\l_*$ get suppressed by a factor 
$e^\tau$ and new series of peaks\footnote{These new peaks are caused not by 
       acoustic oscillations, but by a projection effect: they are peaks in 
       the Bessel function that accounts for free streaming, converting 
       local monopoles at recombination to local quadrupoles at reionization.
       }
are generated starting at the scale $\l_*$. \Fig{fig1} (Left, top panel) 
illustrates that although these new peaks are almost undetectable in $T$, 
drowning in sample variance from the unpolarized Sachs-Wolfe effect, they 
are clearly visible in $E$ and $X$ since the Sachs-Wolfe nuisance is unpolarized 
and absent. The models in \fig{fig1}  have abrupt reionization 
giving $\tau\propto z_*^{3/2}$, so higher $z_*$ is seen to shift the new peaks 
both up and to the right.
For more details about CMB polarization and reionization see \cite{Z97,Hu00}.


\subsubsection{Primordial perturbations:}

As seen in \Fig{fig1} (Right, top panel), gravity waves (a.k.a. tensor fluctuations) 
contribute only to fairly large angular scales, producing $E$ and $B$-polarization. 
Just as for the reionization case, unpolarized fluctuations are also produced but 
are difficult to detect since they get swamped by the Sachs-Wolfe effect. As has 
been frequently pointed out in the literature, no other physical effects (except 
CMB lensing and foregrounds) should produce $B$-polarization, potentially making 
this a smoking gun signal of gravity waves.
Gravitational waves created by inflation would produce $B$-modes in the CMB. 
Because such waves decay after entering the horizon, the spectrum of such 
$B$-mode signal should peak at large angular scales, with an amplitude that 
is tied to the inflationary energy scale.

Adding a small gravity wave component is seen to suppress the correlation $r_\l$ 
in \Fig{fig1} (Right, lower panel), since this component is uncorrelated with 
the dominant signal that was there previously. Indeed, this large-scale correlation 
suppression may prove to be a smoking gun signature of gravity waves that is easier 
to observe in practice than the often-discussed $B$-signal \cite{Knox,Kesden}. This 
$TE$-correlation suppression comes mainly from $E$, not $T$: 
since the tensor polarization has a redder slope than the scalar polarization, 
it can dominate $E$ at low $\l $ even while remaining subdominant in $T$.
Foreground and lensing signals would need to be accurately quantified for
this test, since they would also reduce the correlation.

The amplitudes $\As$, $\At$ and tilts $\ns$, $\nt$ of primordial scalar and tensor 
fluctuations simply change the amplitudes and slopes of the various power spectra:
B is controlled by $(\At,\nt)$ alone, whereas $T$ and $E$ are affected by $(\As,\ns)$ 
and $(\At,\nt)$ in combination. Note that if there are no gravity waves ($\At=0$), 
then these amplitudes and tilts cancel out, leaving the correlation spectrum 
$r_\l$ independent of both $\As$ and (apart from aliasing effects) $n_s$.


\section{Polarization measurements and upper limits}\label{measurements}

\tab
Since the detection of the CMB by Penzias and Wilson in 1965 \cite{PW65}, 
experimentalists have been checking (among other things) if the CMB is also 
polarized. A fact unknown to many is that the first constrain on CMB polarization 
can also be credited to Penzias and Wilson. In their groundbreaking paper, they 
stated that the new radiation they had detected was not only isotropic but also 
unpolarized within the limits of their observations. 
Over the next 37 years, dedicated polarimeters were constructed to set much 
more stringent upper limits on the CMB, culminating in 2002 with its detection 
by the DASI experiment and later re-confirmed by WMAP. These are the
only CMB polarization detections we have so far.
 
DASI was a ground-based experiment located near the Amundsen-Scott South Pole 
Research Station. Observations in all four Stokes parameters were obtained 
within two $3\deg.4$ FWHM fields separated by one hour in RA. The $E$-polarization 
mode was detected at 4.9$\sigma$, while the $X$ cross-polarization mode was 
detected at 2$\sigma$ \cite{kovac02}.
WMAP is an ongoing space mission that produces full sky $I$, $Q$ and $U$ CMB
maps at 5 frequencies between 23 and 94 GHz and angular resolutions from 0$\deg$.82 to 
0$\deg$.21, probing 2 $\simlt \l \simlt$ 600 \cite{bennett03}. The first year 
data have resulted in confirmation of the large scale $X$ cross-correlation at the
10$\sigma$ level, an extraordinary direct evidence of significant reionization 
at higher redshifts. The data agrees with the concordance $\Lambda$CDM model, with 
the best-fit value $\tau$=0.17$\pm$0.04 at 68\% confidence. This implies 
z$_r$=17$\pm$3 \cite{kogut03,page03}.

\begin{figure}[tb]
\preskip
\centerline{\epsfxsize=13.5cm\epsffile{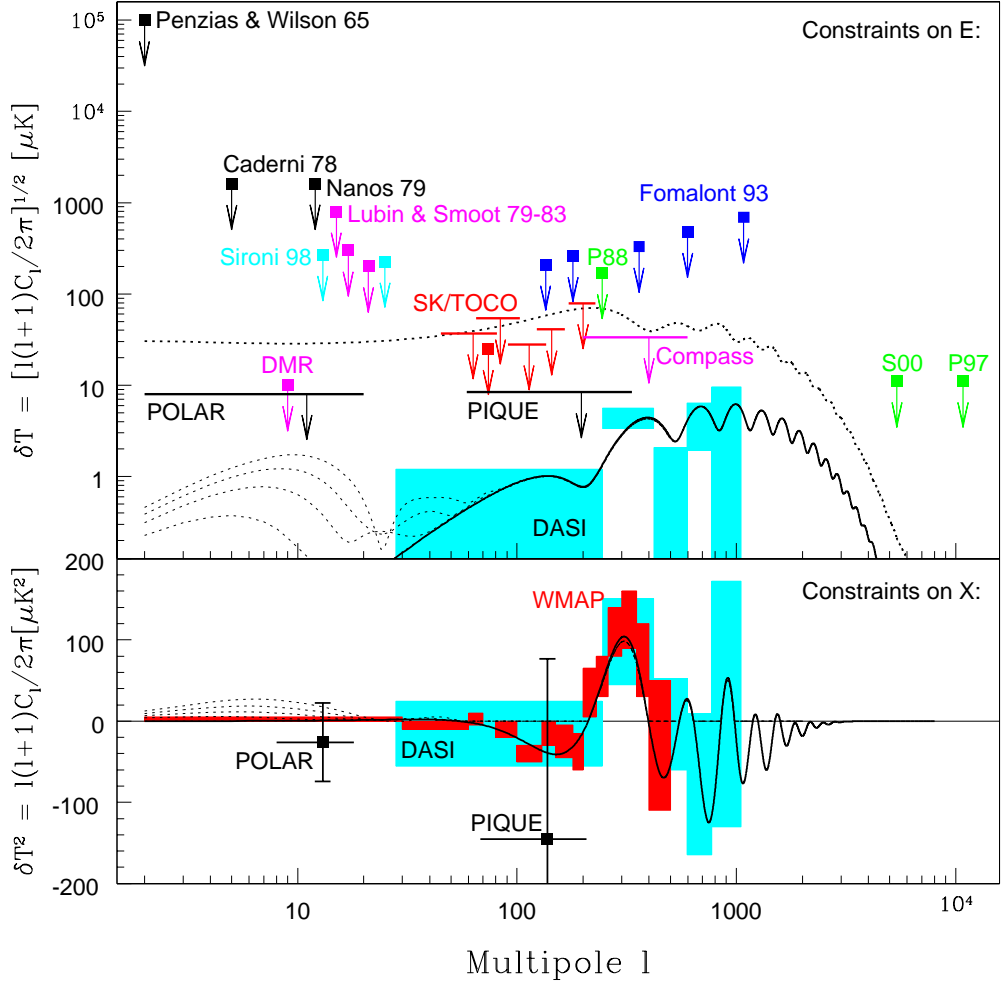}}
\caption{\label{summary}\footnotesize%
 	Summary of constraints on polarization so far. 
	From top to bottom, the three curves show the concordance 
	model predictions for $C_\l^T$, $C_\l^E$ and $C_\l^X$, 
	respectively. Four reionization models with $\tau$=0.1, 0.2, 
	0.3 and 0.4 are also plotted (left dotted lines from bottom 
	to top in both plots).
	The limits for $E$ are shown in the upper panel:
	        Penzias \& Wilson 65 \protect\cite{PW65},
	        Caderni 78 \protect\cite{C78},
	        Nanos   79 \protect\cite{N79},
	        Lubin \& Smoot 79-83 \protect\cite{LS79,LS81,L83} (magenta squares),
	        Sironi 98 \protect\cite{S98} (cyan squares),
		Fomalont 93 \protect\cite{F93},
	        P88 \protect\cite{P88},
		SK/TOCO \protect\cite{W93,T99} (square: SK),
	        S00 \protect\cite{S00},
	        P97 \protect\cite{P97},
		DMR \protect\cite{smootDMR},
		POLAR \protect\cite{keating01} and
 	        PIQUE \protect\cite{H02}, 
		COMPASS \protect\cite{farese03}.
	The limits for $X$ are shown in the lower panel:
	        POLAR \protect\cite{angel_polar} and
	        PIQUE \protect\cite{angel_pique}.
	The shaded cyan and red regions are the DASI \protect\cite{kovac02} 
	and WMAP \protect\cite{page03} results, respectively.
 	}
\end{figure}

We next briefly review the history of CMB polarization measurements (subdividing
them by angular scale). For comparison, we show all measurements and limits on CMB 
polarization to date in \fig{summary}, and a list with some of the ongoing 
and future CMB polarization experiments in \tabela{tabdetectors}


\begin{table}[!hr]
\caption{Ongoing \& \textcolor{blue}{Future} CMB Polarization Experiments}
\medskip
\centerline{\label{tabdetectors}
\begin{tabular}{lccccccr}
\hline
\hline
\multicolumn{1}{l}{Experiment}& 
\multicolumn{1}{c}{FWHM}& 
\multicolumn{1}{c}{$\nu$}&
\multicolumn{1}{c}{Receiver$^a$}& 
\multicolumn{1}{c}{Sensitivity}& 
\multicolumn{1}{c}{Area}& 
\multicolumn{1}{c}{Site-yr$^b$}&
\multicolumn{1}{r}{Ref.}\\
&&[GHz]&&[$\mu$K$\sqrt{s}$]&&&\\
  \hline 
  \hline
  CAPMAP      &4'	       &30,90	       &HEMT	       &1000  &$\delta >$89$\deg$	    &NJ 	   &\protect\cite{barkats03}  \\
  CBI 	      &45'	       &30	       &HEMT	       &      &40$(\deg)^2$		    &Atacama	   &\protect\cite{pearson03}  \\
  DASI 	      &  	       &26-36	       &	       &      &2  $3\deg.4$ Fields	    &SP 	   &\protect\cite{kovac02}    \\
  KuPID       &0$\deg$.2       &15	       & 	       &370   &$\delta >$87$\deg$	    &NJ 	   &\protect\cite{gundersen03}\\
  Polatron    &2.5'	       &100	       &B	       &8000  &Ring/NCP 		    &OVRO	   &\protect\cite{philhour01} \\
  \textcolor{blue}{AMiBA}      &1'-19'  	 &90		&HEMT           &7$\mu$k/hr&Fields:100(')$^2$	     &HI-04         &\protect\cite{lo03}     \\
  \textcolor{blue}{BICEP}      &1$\deg$,0$\deg$.7&100,150	&B		&280   &-5$\deg > \delta >$-25$\deg$ &SP-05 	    &\protect\cite{keating03}\\
  \textcolor{blue}{Polarbear}  &		 &90-350	&B       	&	     & 		    	     &CA-05         &\protect\cite{tran03}     \\
  \textcolor{blue}{QUEST}      &4'  		 &100,150	&B 		&300   &		             &SP-05 	    &\protect\cite{church03} \\
  \textcolor{blue}{SPT}        &     		 & 		&B		&  	     & 		             &SP-06 	    &\protect\cite{carlstrom03}\\
  \hline 
  Archeops    &12'		&143-545	&B		&200   	     &30$\%$ of Sky	     &             &\protect\cite{hamilton04}\\
  B2K         &9.5',6.5',7'	&145,245,345	&B      	&160,290,660 &1284$(\deg)^2$	     &SP           &\protect\cite{montroy03} \\
  MAXIPOL     &10'	        &140,420 	&B      	&130	     &2$\deg$ "bow tie"      &NM    	   &\protect\cite{johnson03} \\
  \hline 
  WMAP            &0$\deg$.82-0$\deg$.21            
                                                &23-94          &               &                      &All Sky      &L2           &\protect\cite{bennett03} \\
  \textcolor{blue}{PlanckLFI}   &33',24',14'	&30,44,70	&HEMT	        &7.7$^*$,10$^*$,18$^*$ &All Sky	     &L2-07   	   &\protect\cite{lawrence03}\\
  \textcolor{blue}{PlanckHFI}   &9.2',7.1' 	&100,143	&B		&...,11$^*$            &All Sky	     &L2-07   	   &\protect\cite{lamarre03} \\
                                &5' 		&217,353 	&B 		&27$^*$,81$^*$         & 	     & 	           &		    \\
                                &5'  		&545,857	&B 		&                      & 	     & 	           &		    \\
  \textcolor{blue}{SPOrt}       &7$\deg$        &22,32,60,90	&HEMT    	&1000                  &80$\%$ of Sky&Sp.Stat.	   &\protect\cite{carretti02}\\
  \hline
  \hline
\end{tabular} 
}
\smallskip
\noindent{\small $^a$B = Bolometer.\\
		            $^b$SP = South Pole;  L2 = An orbit about the 2$^{nd}$ Lagrange point of the Sun-Earth system; 
		            Sp.Stat = Space Station.}\\
\smallskip
\noindent{\small $^*$Sensitivity values are in $\mu$K averaged over the sky for 12 months of integration.}\\
\end{table}


\subsection{Large angular scale (2 $\simlt \l \simlt$ 30)}

\tab
Since the pioneering work of Caderni \cite{C78}, Nanos \cite{N79}, 
Lubin \& Smoot \cite{LS79,LS81,L83} and Sironi \cite{S98}, 
years passed until polarized detector technology achieved sensitivity 
levels that were below the levels of the CMB temperature anisotropy. The first 
of such achievements came on large scales with the POLAR (Polarization 
Observations of Large Angular Regions) experiment  \cite{keating01}. 
POLAR was a ground-based experiment that operated near Madison, Wisconsin. 
It used a simple drift-scan strategy, with a 7$\deg$ FWHM beam at 30 GHz, 
and simultaneously observed the Stokes parameters $Q$ and $U$ in a ring of 
declination $\delta = 43\deg$. The POLAR experiment provided an upper 
limit on $E$- and $B$-modes of 10$\mu$K at 95\% confidence for the multipole 
range of 2 $\simlt \l \simlt$ 20 \cite{keating01}, and an upper limit on the
$X$-mode of 11.1$\mu$K at 95\% confidence level over a similar multipole range
from its cross-correlation with the COBE/DMR map \cite{angel_polar}.
POLAR was later reconfigured to become the COMPASS experiment at intermediate 
angular scales \cite{farese03}. To date, the only CMB polarization detection on
large angular scales is the cross-correlation detected by WMAP.


\subsection{Intermediate angular scale (50 $\simlt \l \simlt$ 1000)}

\tab  
The first upper limits at intermediate angular scales came from the works of
\cite{P88,F93,W93,T99}. 
However, the best upper limit over this same angular range (before its detection
by DASI) was set more than ten years later by PIQUE (Princeton I, Q and U Experiment), 
see \cite{H00}. PIQUE was a CMB polarization experiment on the roof of the physics 
building at Princeton University. It used a single 90 GHz correlation polarimeter 
with FWHM angular resolution of 0$\deg$.235, and observed $Q$ and $U$ in 
a ring of radius of 1$\deg$ around the NCP (North Celestial Pole). 
PIQUE provided an upper limit on $E$- and $B$-modes of 8.4$\mu$K at 95\% 
confidence over the multipole range 59 $\simlt \l \simlt$ 334 \cite{H02}, 
and an upper limit on the $X$-mode of 17.3$\mu$K at 95\% confidence over a 
similar multipole range from its cross-correlation with the SK (Saskatoon) 
map \cite{angel_pique}.
PIQUE has been reconfigured to become the CAPMAP (Cosmic Anisotropy Polarization 
Mapper) experiment \cite{barkats03}, which plans to observe a cap of 1$\deg$ 
radius around the NCP with 4' FWHM at 30 and 90 GHz. 
CAPMAP also shares observing location and technology with KuPID (Ku-band 
Polarization Identifier) \cite{gundersen03}.  KuPID will measure $Q$ and 
$U$ Stokes parameters in a region near the NCP ($\delta > 87\deg$) at 12-18 GHz. 
The primary research objectives of KuPID include surveying the polarized component 
of Galactic synchrotron, characterizing Foreground-X, measuring CMB polarization 
(if foregrounds are not too limiting), and performing follow-up measurements of 
interesting regions identified by WMAP. 

Two other ground-based experiments are being developed to be deployed at the 
South Pole by 2005: BICEP and QUEST. The BICEP (Background Imaging of 
Cosmic Extragalactic Polarization) experiment  \cite{keating03}  will observe 
the South Celestial Pole at 100 and 150 GHz, probing 10 $\simlt \l \simlt$ 200. 
QUEST (Q and U Extragalactic Survey Telescope) is expected to be mounted on 
the DASI teslescope  \cite{church03}. It will also observe the CMB at the same 
frequency range of BICEP, but with a higher angular resolution of FWHM=4'.  

\clearpage
\clearpage

There are also CMB polarization measurements at intermediate scales done from 
balloons. The current generation of balloon-borne experiments include BOOMERanG 
\cite{montroy03}, MAXIPOL \cite{johnson03} and Archeops \cite{benoit03}. 
Polarized BOOMERanG (also known as B2K) made a successful long-duration balloon 
flight over Antarctica during the Austral summer of 2003. It operated at 
145, 245 and 345 GHz with a 10' beam, mapping two regions in the sky: the first 
centered at (RA,DEC) $\approx 75,-45$ with an area of 1161 deg$^2$  and another 
close to the Galactic Plane with an area of 393 deg$^2$.
MAXIPOL had successful flight form Ft. Sumner (NM) in May, 2003. It 
operated at 140 and 420 GHz with an angular resolution of 10'.
Finally, the Archeops experiment also made a successful balloon flight in 2002
and produced maps with measurements of the Galactic dust polarization 
\cite{benoit03}, which we discuss in more details in the next section.


\subsection{Small angular scales ($\l >$ 1000)}

\tab
CMB polarization measurements have also been pursued on smaller scales,
resulting only in upper limits \cite{P97,S00}.
Today, small-scale ground-based experiments, such as AMiBA (Array for 
Microwave Background Anisotropy; \cite{lo03}), SPT (South Pole Telescope;
\cite{carlstrom03}) and ACT (Atacama Cosmology Telescope; \cite{kosowsky03}), 
are working on filling this gap.
AMiBA will operate around 90 GHz and observe the four Stokes parameters.
It is to be deployed on Mauna Kea, Hawaii, with initial observations targeting
$E$-modes by 2004.
SPT is expected to be deployed by 2006 and ACT is planned to be deployed
in Chile, which could also be equipped with a polarimeter.

Also on the works are the next generation of CMB satellites. The Planck 
Surveyor \cite{lawrence03,lamarre03}, which is a dedicated CMB satellite, is scheduled 
to launch in 2007. It will measure the entire sky in 9 frequencies between 
30 and 857 GHz with an angular resolution that can probe 2$\simlt \l \simlt$2000. 
However, the most ambitious plan on the horizon is for a dedicated CMB 
polarization satellite to conduct a search (only foreground limited) for 
signatures of inflationary gravitational waves in the CMB, or to measure 
its $B$-modes. This is the goal of the Inflation Probe in NASA's ``Beyond
Einstein program".


\section{Polarized foregrounds}\label{foregrounds}

\tab
Understanding the physical origin of Galactic microwave emission 
is interesting for two reasons: to determine the fundamental 
properties of the Galactic components, and to refine the modeling 
of foreground emission in CMB experiments. 
At microwave frequencies, three physical mechanisms are known to 
cause foreground contamination: synchrotron and  free-free emission 
(both major contaminants at frequencies below  60 GHz), and dust 
emission (which is a major contaminant above 100 GHz).
When coming from extragalactic objects, this radiation is usually 
referred to as point source contamination and affects mainly small
angular scales. When coming from the Milky Way, this diffuse Galactic 
emission fluctuates mainly on large angular scales. Except for 
free-free emission, all the above mechanisms are known to emit 
polarized radiation. 

Most of the useful information about inflationary gravitational waves 
and reionization is on large angular scales where Galactic foreground 
contamination is the worst, so a key challenge is to model, quantify 
and remove polarized foregrounds.
Unfortunately, these large scales are also the ones where polarized 
foreground contamination is likely to be most severe, both because of 
the red power spectra of diffuse Galactic synchrotron and dust
emission and because they require using a large fraction of the sky, 
including less clean patches. A key challenge in the CMB polarization 
endeavor will therefore be modeling, quantifying and removing large-scale 
polarized Galactic foregrounds.

Studies of CMB polarization must also deal with a second type of foreground,
related to gravitational lensing. Since the deflection of light rays by weak 
gravitational lensing can rotate polarization vectors, CMB $E$-modes can 
be partially converted into $B$-modes with a power  that is proportional to
the lensing signal (see, \eg, \cite{ZSLENS}). Fortunately, such a $B$-component
can be at least partially reconstructed and removed from the CMB using the fact 
that it introduces non-Gaussianities in the data -- see, \eg, \cite{smith03} and 
references therein.


\subsection{Galactic synchrotron emission}

\tab
Unfortunately, we still know basically nothing about the polarized 
contribution of the Galactic synchrotron component at CMB frequencies
(see, \eg,  \cite{allforegpars} and references therein), 
since it has only been measured at lower frequencies and extrapolation 
is complicated by Faraday Rotation. This is in stark contrast to the 
CMB itself, where the expected polarized power spectra and their dependence 
on cosmological parameters has been computed from first principles to high 
accuracy \cite{K97,ZS97,Z98,HuWhite97}. 

There is a recent study of the Leiden surveys \cite{BS76,S84} that try to
shed some light on the properties of the Galactic polarized synchrotron emission 
at the CMB frequencies \cite{angel_polar}.
This study observed that the synchrotron $E$- and $B$-contributions 
are equal to within 10\% from 408 to 820~MHz, with a hint of $E$-domination at 
higher frequencies. One interpretation of this result is that $E>B$ at CMB 
frequencies but that Faraday Rotation mixes the two at low frequencies.
It was also found that Faraday Rotation \& Depolarization effects depend not 
only on frequency but also on angular scale, \ie, they are important at low 
frequencies ($\nu\simlt 10$ GHz) and on large angular scales.
Finally, combining the POLAR and radio frequency results from the Leiden 
surveys, and using the fact that the $E$-polarization of the abundant Haslam signal in 
the POLAR region is not detected at 30 GHz, suggests that the synchrotron polarization 
percentage at CMB frequencies is rather low.
  
In the near future, the best measurement of large-scale Galactic polarized
synchrotron will come from the WMAP satellite. In WMAP's frequency range
(22-90 GHz), the study of its $E$ maps will allow better quantification of
synchrotron, and certainly confirm (or refute) the findings described above.


\subsection{Galactic dust emission}

\tab
Polarized microwave emission from dust is an important foreground that may
strongly contaminate polarized CMB studies unless accounted for. At higher 
frequencies ($\simgt$ 100 GHz) the main contamination comes from vibrational 
dust emission, while at lower frequencies (15 $\simlt \nu \simlt$ 60 GHz) it may 
come from another dust population composed basically of small grains  
that emit radiation via rotational rather than vibrational excitations \cite{DL98}.

This small grain component, nicknamed Foreground-X \cite{tenerife2}, is spatially 
correlated with the 100 $\mu$m dust emission but with a spectrum that rises towards 
lower frequencies, subsequently flattening and turning down somewhere around 
15 GHz.  Although there is plenty of observational evidence in favor of its existency
(see \cite{tenerife3} and references therein), 
there is no spatial template for this component. If Foreground-X is due to spinning 
dust particles, the amount of polarization of this component is marginal for $\nu \simgt$ 35 GHz.
However, if Foreground-X emission is due to the magneto-dipole mechanism the 
polarization can be substantial -- see \cite{LF03} for details.

For the case of polarized vibrational dust emission, little is known, and experiments 
such as Archeops and B2K are probably our best short-term hope for trying to understand 
its behavior at microwave frequencies.
For instance, the Archeops experiment detected polarized emission by dust at 
353 GHz \cite{benoit03}. They find that  the diffuse emission from the Galactic plane 
is 4-5\% polarized,  and its orientation is mostly perpendicular to the plane. There is 
evidence for a powerful grain alignment mechanism throughout the interstellar medium. 
 

\section{Conclusions: what to expect for the future?}\label{conclusions}

\tab
CMB polarization is likely to be a goldmine of cosmological information, allowing to 
improve measurements of many cosmological parameters and numerous important 
cross-checks and tests of the underlying theory.   
For some of the future goals, such as to detect gravity waves through CMB 
polarization, we will need to develop new polarized detector technology and 
better understand the polarized foregrounds (Galactic and extragalactic). In fact, 
our ability to measure cosmological parameters using the CMB will only be as good 
as our understanding of the microwave foregrounds.  
To do a good job removing foregrounds, we need to understand their frequency 
and scale dependence, frequency coherence, and better characterize their non-Gaussian 
behavior.

\smallskip
\noindent
{\bf  Acknowledgements:} The author wish to thank Max Tegmark for proof-reading the manuscript. 
Support was provided by the NASA grant NAG5-11099.
  

\end{document}